\documentclass[reprint,superscriptaddress,amsmath,amssymb,aps,prl,floatfix]{revtex4-2}
\usepackage{graphicx}
\usepackage{bm}
\usepackage{hyperref}
\usepackage{todonotes}
\usepackage{nicefrac}
\usepackage{upgreek}
\usepackage[normalem]{ulem}

\definecolor{py_color}{cmyk}{1, 0.3, 0, 0}
\definecolor{cwh_color}{cmyk}{0, 0.8, 0.8, 0}
\definecolor{hn_color}{cmyk}{0.8, 0, 0.8, 0}

\begin{document}
\newcommand{\SRO}{Sr\textsubscript{2}RuO\textsubscript{4}}

\title{Probing momentum-dependent scattering in uniaxially stressed \SRO{} through the Hall effect}

\author{Po-Ya Yang}
\affiliation{Max Planck Institute for Chemical Physics of Solids, N\"{o}thnitzer Str 40, 01187 Dresden, Germany}
\author{Hilary M. L. Noad}
\affiliation{Max Planck Institute for Chemical Physics of Solids, N\"{o}thnitzer Str 40, 01187 Dresden, Germany}
\author{Mark E. Barber}
\affiliation{Max Planck Institute for Chemical Physics of Solids, N\"{o}thnitzer Str 40, 01187 Dresden, Germany}
\affiliation{Department of Applied Physics, Stanford University, Stanford, CA 94305, USA}
\affiliation{Geballe Laboratory for Advanced Materials, Stanford, CA 94305, USA}
\author{Naoki Kikugawa}
\affiliation{National Institute for Materials Science, Tsukuba 305-0003, Japan}
\author{Dmitry Sokolov}
\affiliation{Max Planck Institute for Chemical Physics of Solids, N\"{o}thnitzer Str 40, 01187 Dresden, Germany}
\author{Andrew P. Mackenzie}
\affiliation{Max Planck Institute for Chemical Physics of Solids, N\"{o}thnitzer Str 40, 01187 Dresden, Germany}
\affiliation{Scottish Universities Physics Alliance, School of Physics and Astronomy, University of St. Andrews, St. Andrews KY16 9SS, U.K.}
\author{Clifford W. Hicks}
\affiliation{School of Physics and Astronomy, University of Birmingham, Birmingham B15 2TT, U.K.}
\affiliation{Max Planck Institute for Chemical Physics of Solids, N\"{o}thnitzer Str 40, 01187 Dresden, Germany}

\date{\today}

\begin{abstract}

Under in-plane uniaxial stress, the largest Fermi surface sheet of the correlated metal \SRO{} undergoes a Lifshitz transition from an electron-like to an open geometry. We
investigate the effects of this transition on transport through measurement of the longitudinal resistivity $\rho_{xx}$ and the Hall coefficient $R_\text{H}$. At temperatures where
scattering is dominated by electron-electron scattering, $R_\text{H}$ becomes more negative across the Lifshitz transition, opposite to expectations from the change in Fermi
surface topology. We show that this change in $R_\text{H}$ is explainable only if scattering changes throughout the Brillouin zone, not just at the point in $k$-space where the
Lifshitz transition occurs. In a model of orbital-dependent scattering, the electron-electron scattering rate on sections of Fermi surface with $xy$ orbital weight decreases
dramatically. On the other hand, at temperatures where defect scattering dominates $\rho_{xx}$ and $R_\text{H}$ are essentially constant across the Lifshitz transition. 

\end{abstract}

\maketitle

\textit{Introduction.} The Hall coefficient $R_\text{H}$ of multi-band metals is a challenging quantity to analyze. In the low-field limit, it is determined by an integral of the
mean free path around the Fermi surfaces~\cite{Ong91_PRB}. It can therefore be used to probe the momentum dependence of scattering in metals with simple Fermi
surfaces~\cite{Nandi18_npj, Narduzzo08_PRB, French09_NJP}. However, as complexity increases models become badly underconstrained. The case of the correlated metal and superconductor
\SRO{}~\cite{Maeno94_Nature} highlights the challenge. With 1\% substitution of La onto the Sr site, the low-temperature Hall coefficient changes from electron-like to hole-like,
even though the Fermi surfaces are almost unchanged~\cite{Kikugawa04_PRB1, Kikugawa04_PRB3}.

The advent of strain tuning might make $R_\text{H}$ a more broadly useful measurement quantity. By correlating changes in $R_\text{H}$ with specific strain-driven changes in
electronic structure, models of scattering can be tested with greater rigor. Crucially, if the deformation is elastic the defect landscape is not altered. 

Here, we study \SRO{}, which has become an important test case for understanding of correlated electron materials~\cite{Kugler20_PRL, Mravlje11_PRL, Deng19_NatComm}. It is a highly
two-dimensional metal, in which correlations renormalize but do not destroy Landau quasiparticles~\cite{Bergemann00_PRL, Bergemann03_AP}. Under uniaxial stress $\sigma$ applied
along the $[100]$ lattice direction, the largest Fermi surface sheet [the $\gamma$ sheet--- see Fig.~1(a)] undergoes a Lifshitz transition from an electron-like to an open
geometry~\cite{Sunko19_npj}. This transition is illustrated in Fig.~1(b). It has been reported to occur at $\sigma = \sigma_\text{L} = -0.71 \pm 0.08$~GPa~\cite{Barber19_PRB},
where negative values denote compression. Under the common assumption that the Hall effect is determined by the topology of the Fermi surfaces, this transition should make
$R_\text{H}$ more positive. If scattering time rather than mean free path is isotropic, the quantitative change in $R_\text{H}$ might be small, but is still expected to be
positive~\cite{Maharaj17_PRB, Kokkinis22_PRB}.

\begin{figure}[ptb]
\includegraphics[width=85mm]{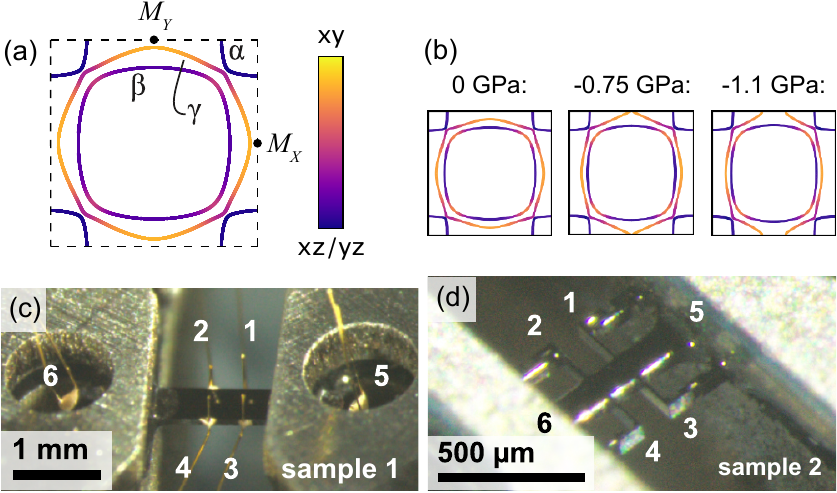}
\caption{(a) Model Fermi surfaces of \SRO{}, coloured by orbital weight, from Ref.~\cite{Zingl19_npj}. (b) Fermi surfaces calculated with a simple tight-binding model, presented
later in this Letter, under uniaxial stresses along $[100]$ of $0$, $-0.75$, and $-1.1$~GPa.  (c,d) Photographs of samples 1 and 2.}
\end{figure}

We report here that in fact the Lifshitz transition makes $R_\text{H}$ more \textit{negative}, opposite to the expectation from the change in Fermi surface topology. Furthermore,
the magnitude of the change implies changes in scattering throughout the Brillouin zone: going through the Lifshitz transition appears to make \SRO{} a considerably less-strongly
correlated metal. \SRO{} therefore provides an important example of how electronic correlations emerge in metals, and how correlations may potentially be ``switched on and off.''

\textit{Methods.} Measurements were performed using a piezoelectric-driven uniaxial stress device similar to that described in Ref.~\cite{Barber19_RSI}, that incorporates sensors
of both the applied displacement and applied force. As in Ref.~\cite{Jerzembeck22_NatComm}, samples were mounted onto carriers, which were then mounted onto the cell.  These
carriers comprise two parts, and the sample is compressed when they are brought together. This mechanism allows in situ calibration of the zero-force reading of the force sensor.

Two samples, photographed in Figs.~1(c,d), were studied. Sample 1 is a beam of constant cross-section with electrical contacts attached by hand, while for sample 2 a plasma focused
ion beam was used to mill contacts from the sample itself. The residual resistivities are 0.4 and 0.09~$\mu\Omega$-cm, respectively. Due to its lower residual resistivity, most of
the data presented here are from sample 2, though some data from sample 1 are included to demonstrate reproducibility. For both, the Hall voltage $V_\text{H}$ at field $B$ is taken
as $\frac{1}{2}[V_{13}(B) - V_{13}(-B)]$ or $\frac{1}{2}[V_{24}(B) - V_{24}(-B)]$, where $V_{ij} \equiv V_i - V_j$, $V_i$ is the voltage in contact $i$, and contact numbering is
shown in Figs.~1(c,d). This procedure cancels contributions to $V_H$ from contact misalignment. The Hall coefficient $R_\text{H}$ is then obtained as $R_\text{H} =
V_\text{H}t/(IB)$, where $I$ is the applied current and $t$ the sample thickness.

\begin{figure}[ptb]
\includegraphics[width=85mm]{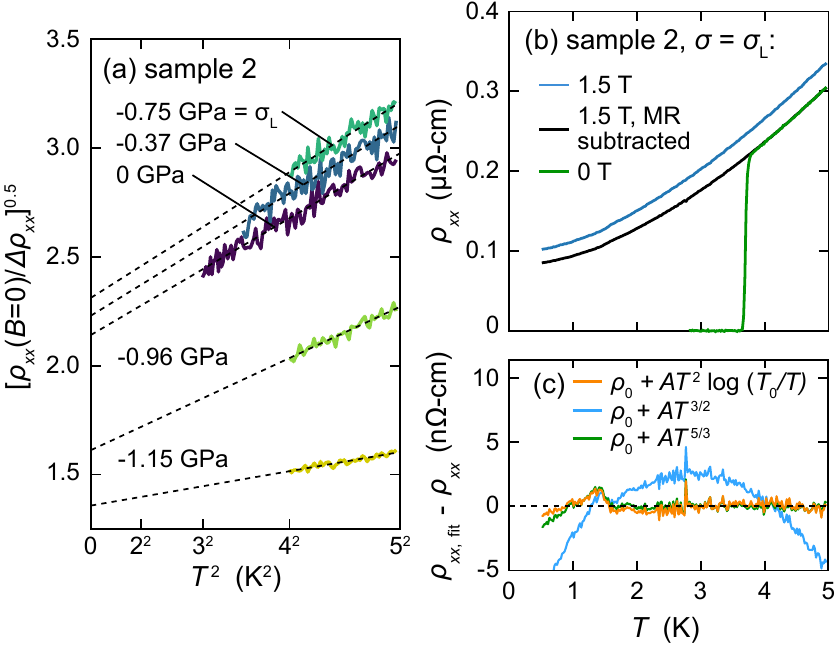}
\caption{(a) $[\rho_{xx}(B=0) / \Delta \rho_{xx}]^{0.5}$, where $\Delta \rho_{xx}$ is the change in $\rho_{xx}$ between 0 and 1.5~T, against $T^2$ for sample 2 at various stresses.
The dashed lines are linear fits. (b) $\rho_{xx}(T)$ of sample 2 at $\sigma = \sigma_\text{L} = -0.75$~GPa. The black line is an average of
data at $+1.5$ and $-1.5$~T, with the magnetoresistance subtracted as described in the text. (c) The difference between various fitting models and the black line in panel (b).}
\end{figure}

\textit{Longitudinal resistivity.} We look first at the temperature dependence of the longitudinal resistivity $\rho_{xx}$ at the Lifshitz transition, testing our data against
three hypothesized temperature dependences. (1) $\rho_{xx} \propto T^2 \log T$. Boltzmann transport theory applied to a Fermi liquid tuned to an electron-to-hole Lifshitz
transition yields a $T^2 \log T$ form~\cite{Hlubina96_PRB, Mousatov20_PNAS}. To obtain this result, it is necessary to take into account the fact that, in the absence of impurity
scattering, only umklapp scattering generates resistivity.  Ref.~\cite{Stangier22_PRB} finds that $\rho_{xx} \propto T^2 \log T$ also for an electron-to-open transition, as occurs
here, but only if there are other bands present. (2) $\rho_{xx} \propto T^{3/2}$. The non-umklapp scattering rate, which affects \textit{e.g.} the thermal but not the electrical
resistivity, is expected to scale as $T^{3/2}$~\cite{Stangier22_PRB}. Also, $\rho_{xx} \propto T^{3/2}$ is expected for extended saddle points~\cite{Hlubina96_PRB}. (3) $\rho_{xx}
\propto T^{5/3}$. This form has been observed near ferromagnetic quantum critical points~\cite{Nicklas99_PRL, Smith08_Nature, Sutherland12_PRB}. It is understood theoretically as a
result of loss of quasiparticle coherence due to the quantum criticality~\cite{Smith08_Nature, Sutherland12_PRB, Mathon68_PRSA}. 

\begin{figure}[ptb]
\includegraphics[width=85mm]{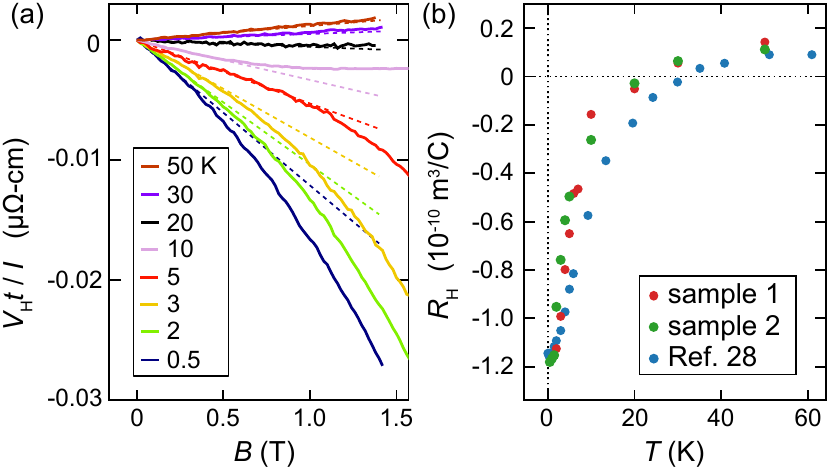}
\caption{(a) $V_\text{H}t/I$, where $t$ is sample thickness and $I$ the applied current, versus $B$ for sample~2 at various temperatures. The dotted lines are extrapolations of
linear fits over $0 < B < 0.35$~T. Equivalent data for Sample 1 are shown in~\cite{SM}. (b) $R_H(T)$ of both samples at zero stress, and $B$ up to $\pm 0.5$~T for
sample 1 and $\pm 0.35$~T for sample 2. Data from Ref.~\cite{Mackenzie96_PRB} are also shown.}
\end{figure}

\begin{figure}[tb]
\includegraphics[width=85mm]{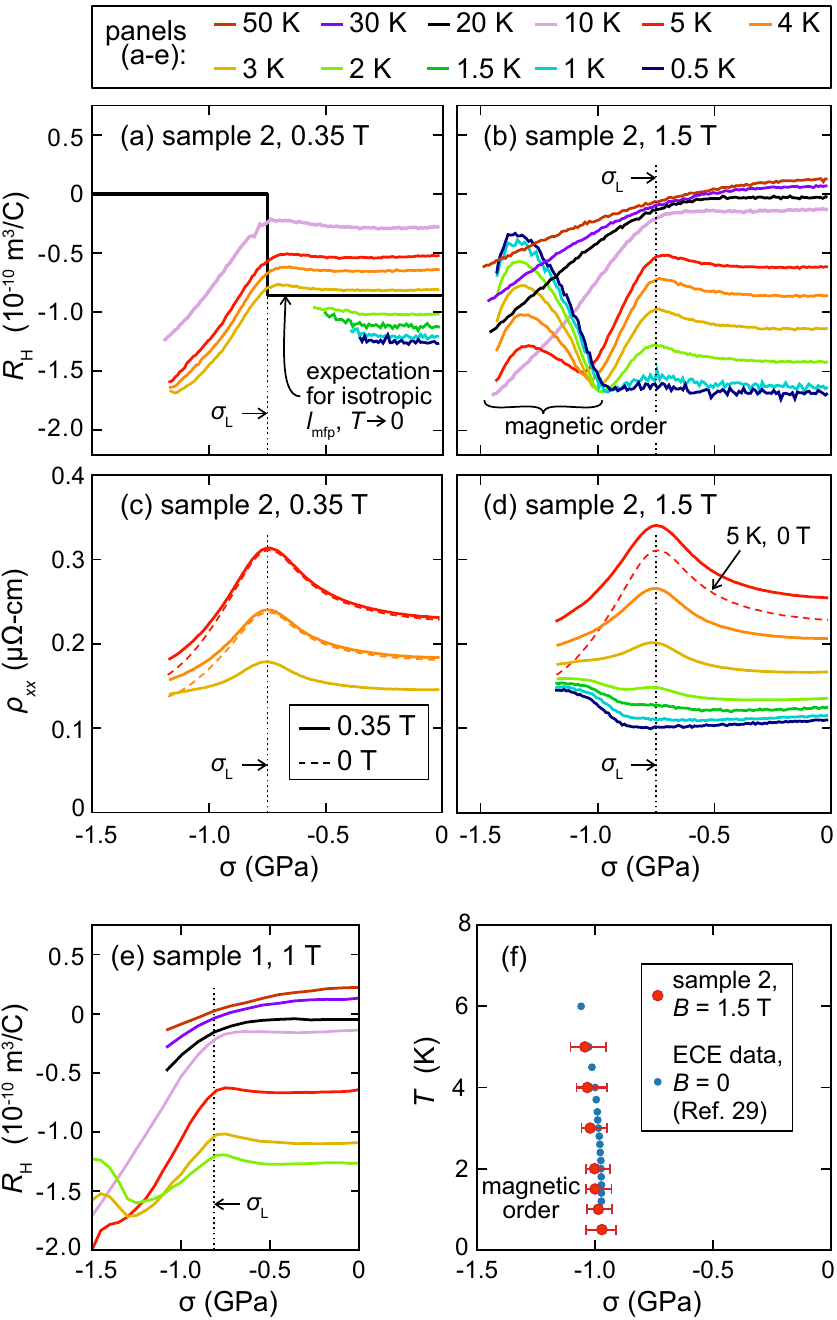}
\caption{(a) $R_\text{H}(\sigma)$ of sample 2 at various temperatures, measured at $B = \pm 0.35$~T. Data that are clearly affected by superconductivity are excluded. The expected
$R_\text{H}(\sigma)$ for isotropic mean free path and $T \rightarrow 0$, calculated using a simple tight-binding model, is also shown. (b) $R_\text{H}(\sigma)$ of sample~2 at $B =
\pm 1.5$~T, a strong enough field to fully suppress the superconductivity. (c) $\rho_{xx}(\sigma)$ of sample~2 at $B = 0$ and $0.35$~T. (d) $\rho_{xx}(\sigma)$ of sample~2 at $B =
1.5$~T. (e) $R_\text{H}(\sigma)$ of sample 1 at $B = \pm 1$~T. (f) The phase boundary of the magnetic order, taken as the stresses where $d^2R_\text{H}/dT^2$ in panel (c) is
maximum. The error bars are the FWHM of the peaks in $d^2 R_\text{H}/dT^2$. Also shown is the phase boundary found in elastocaloric effect data~\cite{Li22_Nature}, with the
Lifshitz stress normalized to $-0.75$~GPa.}
\end{figure}

The superconductivity of \SRO{} is strongly enhanced at the Lifshitz transition~\cite{Steppke17_Science}. To measure $\rho_{xx}$ at low temperatures, we suppress it with a 1.5~T
field applied along the $c$ axis. We apply a magnetoresistance model to subtract an estimate for $\Delta \rho_{xx}$, the change in $\rho_{xx}$ under the 1.5~T field, yielding an
estimate for the zero-field resistivity that would be obtained without superconductivity. For isotropic metals in weak magnetic fields, 
\begin{equation}
\label{eq:MR}
\Delta \rho_{xx} \propto \omega_\text{c}^2 \tau^2 \times \rho_{xx}(B = 0),
\end{equation}
where the cyclotron frequency $\omega_\text{c} = eB/m^*$ and $\tau$ is the relaxation time. $\tau^{-1} = \alpha + \beta T^2$ is expected in the Fermi liquid regime, and in
unstressed \SRO{} it has been shown that $(\rho_{xx}/\Delta \rho_{xx})^{1/2}$ can be fitted by $\alpha + \beta T^2$ up to at least 80~K~\cite{Hussey98_PRB}, even though $\rho_{xx}$
follows a $T^2$ form only up to $\sim 30$~K~\cite{Maeno97_JPSJ, Hussey98_PRB}. In Fig.~2(a) it is shown that a model $(\rho_{xx}/\Delta \rho_{xx})^{1/2} = \alpha + \beta T^2$ fits
the observed magnetoresistance at each stress that we tested. We use this model to extrapolate $\Delta \rho_{xx}$ into the superconducting regime where it cannot be directly
measured.

Fig.~2(b) shows $\rho_{xx}(T)$ and $\rho_{xx}(T) - \Delta \rho_{xx}(T)$ of sample 2 at $\sigma = \sigma_L$. Fig.~2(c) shows the differences between $\rho_{xx} - \Delta \rho_{xx}$
and the three hypothesised fitting models. $\rho_{xx} = \rho_0 + A T^{3/2}$ clearly does not fit the data well, while $\rho_{xx} = \rho_0 + AT^{5/3}$ and $\rho_{xx} = \rho_0 + A
T^2 \log (T_0 / T)$ work equally well. For the latter, we obtain $T_0 = 95$~K. In Ref.~\cite{SM}, we show that an alternative magnetoresistance model, $(\rho_{xx}/\Delta
\rho_{xx})^{1/2} = \alpha + \beta T^{3/2}$, can also be applied at $\sigma = \sigma_\text{L}$, and does not alter the conclusion that $\rho_{xx} = \rho_0 + A T^2 \log (T_0 / T)$ is
a better fit to the data than $\rho_{xx} = \rho_0 + AT^{3/2}$. In contrast, in earlier work it was found that the $T^2 \log T$ and $T^{3/2}$ forms worked equally well at $\sigma =
\sigma_\text{L}$, over a fitting range of 4--40~K~\cite{Barber18_PRL}. We hypothesize that this temperature range was too high to accurately capture the low-temperature behavior.

Although the $T^{5/3}$ and $T^2 \log T$ models work equally well, the hypothesis under which $T^{5/3}$ is obtained theoretically, ferromagnetic quantum criticality, appears not to
apply here. NMR data on \SRO{} show that the Stoner factor of \SRO{} is enhanced by $\sim$30\% at the Lifshitz transition~\cite{Luo19_PRX}, and that there is no dramatic change in
quasiparticle weight~\cite{Chronister21_npj}. Tuning to the Lifshitz transition strengthens ferromagnetic fluctuations, but apparently not to point that they become critical. 

\textit{Hall effect.} We now look at the Hall effect. To investigate appropriate measurement fields, in Fig.~3(a) we show $V_\text{H}$ versus $B$ for sample~2. For temperatures
below $\sim 10$~K, the low-field regime in which $V_\text{H} \propto B$ extends only up to $\sim 0.3$~T.  Fig.~3(b) shows $R_\text{H}(T)$ of both samples in the low-field limit.
$R_\text{H}$ is hole-like at 50~K, and electron-like below $\sim 20$~K, as reported previously~\cite{Shirakawa95_JPSJ, Mackenzie96_PRB}.

Figs.~4(a,c) show $R_\text{H}(\sigma)$ and $\rho_{xx}(\sigma)$ of sample~2 at various temperatures and at $\pm 0.35$~T, which is close to the low-field limit. Figs.~4(b,d) show the
equivalent data at $\pm 1.5$~T, which fully suppresses the superconductivity. Fig.~4(e) shows data on $R_\text{H}$ from sample~1 at 1~T. We identify the Lifshitz stress
$\sigma_\text{L}$ as the peak in $\rho_{xx}$ at 5~K, yielding $\sigma_\text{L} = -0.75 \pm 0.08$~GPa for sample~2. For sample~1, $\rho_{xx}$ was not measured, but comparison of
$R_\text{H}(\sigma)$ with that of sample~2 yields $\sigma_\text{L} = -0.82 \pm 0.08$~GPa. For temperatures of $\sim 2$~K and above, the data shown in Figs.~4(a, b, e) reveal that
$R_\text{H}$ decreases as samples are compressed through $\sigma = \sigma_\text{L}$, that is, moving from right to left in the graphs. Below $\sim 5$~K, this decrease is somewhat
less steep for sample~1 than sample~2, possibly due to the stronger defect scattering, but the difference is not large.  Fig.~4(a) also shows the expected $R_\text{H}(\sigma)$ for
an isotropic mean free path, calculated using a tight-binding model that will be presented, and the contrast with the data is clear. 

In the 1.5~T data in Figs.~4(b,d) it can be seen that below $\sim 1$~K, both $R_\text{H}$ and $\rho_{xx}$ become essentially constant across the Lifshitz transition. In other
words, even in the $T \rightarrow 0$ limit, $R_\text{H}$ does not respond as expected to the change in Fermi surface topology.

Before discussing the Hall data further, we comment briefly on stress-induced magnetic order. This order is probably an incommensurate spin density wave~\cite{Grinenko21_NatPhys},
and has a clear effect on $R_\text{H}$ at 1.5~T. The stress range where the magnetic order condenses is indicated in Fig.~4(b). However, at $0.35$~T the magnetic order has no
clear effect on $R_\text{H}(\sigma)$ [see Fig.~4(a)], and at 1~T $R_\text{H}(\sigma)$ shows some response to the order but less than at 1.5~T [see Fig.~4(e)].  This variation is
not because the magnetic phase boundary has strong field dependence: in Fig.~4(f), we show that the magnetic phase boundary identified in the 1.5~T data matches well that
identified from elastocaloric effect measurements in zero field~\cite{Li22_Nature}. Rather, there must be a strong field dependence in the effect of the magnetic phase on
transport. We leave this as a topic for future investigation.

\textit{Discussion.} We focus first on $R_\text{H}$ at 5--10 K, where the downturn in $R_\text{H}$ at $\sigma < \sigma_\text{L}$ is large and quite sharp.  $\rho_{xx}$ peaks at
$\sigma = \sigma_\text{L}$ due to enhanced electron-electron scattering~\cite{Mousatov20_PNAS, Herman19_PRB}, and the fact that both the downturn in $R_\text{H}$ and the peak in
$\rho_{xx}$ fade over the same temperature range shows that the downturn in $R_\text{H}$ is also associated with electron-electron scattering. In this section, we show that the
magnitude of the change in $R_\text{H}$ implies changes in scattering throughout the Brillouin zone, and that a model of orbital differentiation can account for the change.

Following Ref.~\cite{Zingl19_npj}, in this orbital differentiation model the band-dependent electron-electron scattering rate $\eta_\nu(\mathbf{k})$, where $\nu = \alpha$, $\beta$,
or $\gamma$, is given by:
\begin{equation} 
\eta_\nu(\mathbf{k}) = \sum_m \left|\left \langle \chi_m(\mathbf{k}) | \psi_\nu(\mathbf{k}) \right \rangle \right|^2 \eta_m.  
\label{eq:orbitalDecomposition} 
\end{equation} 
$\alpha$, $\beta$, and $\gamma$ label the Fermi surfaces--- see Fig.~1(a). $\left|\left \langle \chi_m(\mathbf{k}) | \psi_\nu(\mathbf{k}) \right \rangle \right|^2$ is the weight of
orbital $m$ in band $\nu$ at momentum $\mathbf{k}$, and $\eta_m$ is an orbital-dependent scattering rate. Scattering is taken to be a local property, so $\eta_m$ has no
$\mathbf{k}$ dependence. This model is supported by photoemission and Raman scattering data~\cite{Tamai19_PRX, Philippe21_PRB}, and applies when Hund's interactions control
electronic correlations, as appears to be the case in \SRO{}~\cite{Mravlje11_PRL, Deng19_NatComm}.  

We calculate the orbital weights using the simplest possible tight-binding model that reproduces the Fermi surface topology of \SRO{}~\cite{Zabolotnyy2013_JoESaRP, Romer2019_PRL,
Cobo2016_PRB, SM}. Our model Fermi surfaces are shown in Fig.~1(b). Near-neighbour hopping integrals are taken to vary linearly with strain, with a scaling constant
that is set so that the Lifshitz transition occurs at $\sigma_\text{L} = -0.75$~GPa. The Hall conductivity $\sigma_{xy}$ is calculated using the Ong construction~\cite{Ong91_PRB}: 
\begin{equation} 
\sigma_{xy}^{} = \frac{e^3}{2 \pi^2 \hbar^2} \boldsymbol{B} \cdot \int_{FS} \frac{d \mathbf{l}_\text{mfp} (\mathbf{k}) \times \mathbf{l}_\text{mfp} (\mathbf{k})}{2},
\label{eq:sigma_xy} 
\end{equation} 
where $\mathbf{l}_\text{mfp}(\mathbf{k})$ is the mean free path at point $\mathbf{k}$. Stated in words, Eq.~\ref{eq:sigma_xy} states that the Hall conductivity is determined by the
curvature of the sections of Fermi surface with longer mean free path. To obtain $R_\text{H}$, the longitudinal conductivities $\sigma_{xx}$ and $\sigma_{yy}$ are needed:
$R_\text{H} \approx (1/B)\sigma_{xy}/(\sigma_{xx}\sigma_{yy})$. $\sigma_{xx}$ is taken to be: 
\begin{equation}
\sigma_{xx} = \frac{e^2}{2\hbar\pi^2}\int_{FS}ds \: l_\text{mfp}(\mathbf{k})\left(\mathbf{l}_\text{mfp}\cdot\mathbf{\hat{x}}\right)^2.  
\label{eq:longitudinalConductivity}
\end{equation} 
Eq.~\ref{eq:longitudinalConductivity} neglects the distinction between umklapp and non-umklapp processes; it is an approximate model. We also neglect impurity and phonon
scattering, so $l_\text{mfp}(\mathbf{k}) = v_\text{F}(\mathbf{k})/\eta_\nu(\mathbf{k})$, where $v_\text{F}(\mathbf{k})$ is the Fermi velocity. 

\begin{figure}[ptb]
\includegraphics[width=85mm]{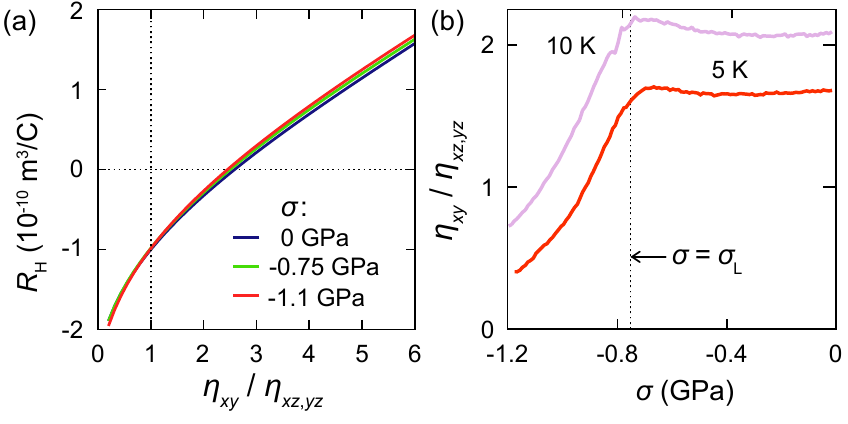}
\caption{(a) $R_\text{H}$ calculated within the orbital-dependent scattering model described in the text, as a function of the ratio of $xy$ and $xz$, $yz$ scattering rates,
$\eta_{xy} / \eta_{xz, yz}$. The calculation employs a simple tight-binding model of the Fermi surfaces of \SRO{}. (b) $\eta_{xy}/\eta_{xz, yz}$ versus stress at 5 and 10~K,
determined from $R_\text{H}$ of sample 2 at $\pm 0.35$~T and the calculation results shown in panel (a).}
\end{figure}

For simplicity, we assume $\eta_{xz} = \eta_{yz} = \eta_{xz,yz}$, even at nonzero stress where this equality is not symmetry-protected.  $R_\text{H}$ calculated as a function of
the ratio $\eta_{xy} / \eta_{xz, yz}$ and at three different stresses is shown in Fig.~5(a). Our results at $\sigma = 0$ match well the more accurate model of
Ref.~\cite{Zingl19_npj}.  $R_\text{H} > 0$ for large $\eta_{xy}/\eta_{xz, yz}$ because both the $\beta$ and $\gamma$ sheets have substantial $xy$ orbital weight. A high $xy$
scattering rate suppresses the contribution to the Hall conductivity from these sheets, leaving that from the $\alpha$ sheet, which is hole-like.

We use the results shown in Fig.~5(a) to convert $R_\text{H}$ measured at $\pm 0.35$~T and at 5 and 10~K, temperatures where electron-electron scattering
dominates~\cite{Deng16_PRL}, to $\eta_{xy}/\eta_{xz, yz}$.  Results are shown in Fig.~5(b). At $\sigma = 0$, we find $\eta_{xy}/\eta_{xz,yz}>1$. This is expected because in
unstressed \SRO{} it is the $xy$ band that is most strongly renormalized, and the electron-electron scattering rate is correlated with the degree of
renomalization~\cite{Deng14_PRL}. Dynamical mean-field theory calculations indicate that this stronger renormalization is due to proximity to the Lifshitz
transition~\cite{Mravlje11_PRL, Kugler20_PRL}, which occurs in the $xy$ band. It is therefore reasonable to hypothesize that tuning even closer to this transition would cause a
further increase in $\eta_{xy} / \eta_{xz, yz}$. Such an increase is indeed visible in Fig.~5(b), but it is small. The much more prominent feature is a steep fall in
$\eta_{xy}/\eta_{xz, yz}$ for $\sigma < \sigma_\text{L}$. Consistent with this interpretation that scattering in the $xy$ band falls steeply for $\sigma < \sigma_\text{L}$, the
magnetoresistance becomes much stronger for $\sigma < \sigma_\text{L}$: in Fig. 4(c), a noticeable difference appears between $\rho_{xx}(T = 5\;\text{K})$ at 0 and 0.35 T.

Even if this specific model is called into question, our data show that scattering must change throughout the Brillouin zone. Our tight-binding model yields $R_\text{H} = -1.74
\times 10^{-10}$~m$^3$/C for unstressed \SRO{} when $l_\text{mfp}$ is isotropic and equal on the $\beta$ and $\gamma$ sheets but zero on the $\alpha$ sheet. The fact that
$R_\text{H}$ falls to values in this range when $\sigma < \sigma_\text{L}$ means that $l_\text{mfp}$ must become long on essentially all convex portions of the $\beta$ and $\gamma$
sheets.

We briefly discuss the $T \rightarrow 0$ limit, where impurity scattering dominates. In this limit it is frequently assumed that $l_\text{mfp}$ is set by defect spacing and is
therefore isotropic. However, we have already noted that with isotropic $l_\text{mfp}$ $R_\text{H}$ would jump upward at the Lifshitz transition, due to the change in Fermi surface
topology. The fact that $R_\text{H}$ remains essentially constant across $\sigma = \sigma_\text{L}$ indicates that $l_\text{mfp}$ must be very short in the neck region of the
$\gamma$ sheet. The observation here that $\rho_{xx} \propto T^2 \log T$ at $\sigma = \sigma_\text{L}$ indicates standard quasiparticle dispersion right up to the saddle point, and
we therefore hypothesize that small-angle scattering from charge disorder is stronger in these regions. Similar anisotropy is seen in overdoped cuprate superconductors, where
$l_\text{mfp}$ is shorter along the $(0,\pi)$ and $(\pi,0)$ directions (the anti-nodal directions)~\cite{Narduzzo08_PRB, French09_NJP, Ayres21_Nature}, and enhanced small-angle
scattering due to charge disorder is thought to play a role~\cite{Narduzzo08_PRB, Abrahams00_PNAS}. 

In summary, going through the Lifshitz transition drives large changes in electron-electron scattering in \SRO{}. In a model of orbital differentiation, the $xy$ band transitions
over a small strain range from being the most- to being the least-strongly corrleated band. This result provide important information on the origin of electronic correlations in
metals, and should be tested with other techniques.

\section{APPENDIX}

\textit{Sample carrier.} The sample carrier used here is illustrated in Fig.~S1(a). The two parts are respectively bolted to the uniaxial stress cell, which controls the distance
between them. Bringing the two parts into contact compresses the sample. The set of four flexures in the sample carrier, which protect the sample from inadvertent transverse and
twisting forces, are calculated to have a combined spring constant of 0.14~N/$\mu$m. This spring constant is used to calculate the force in the flexures when the sample gets
compressed, and is subtracted from the force reading from the cell to obtain the force applied to the sample.
\\

\textit{Sample 2.} As described in the main text, for sample 2 a plasma focused ion beam was used to mill voltage contacts from the sample itself, ensuring highly accurate
placement. In Fig.~S1(b), a scanning electron micrograph of sample 2 is shown after mounting onto the sample carrier, but before the upper plates were installed. The electrical
contacts at the end of the sample are visible in this SEM. In Fig.~S1(c), a photograph taken after the upper plates were installed is shown. The upper plate is held in place with
Stycast 2850.
\\

\begin{figure}[ptb]
\includegraphics[width=85mm]{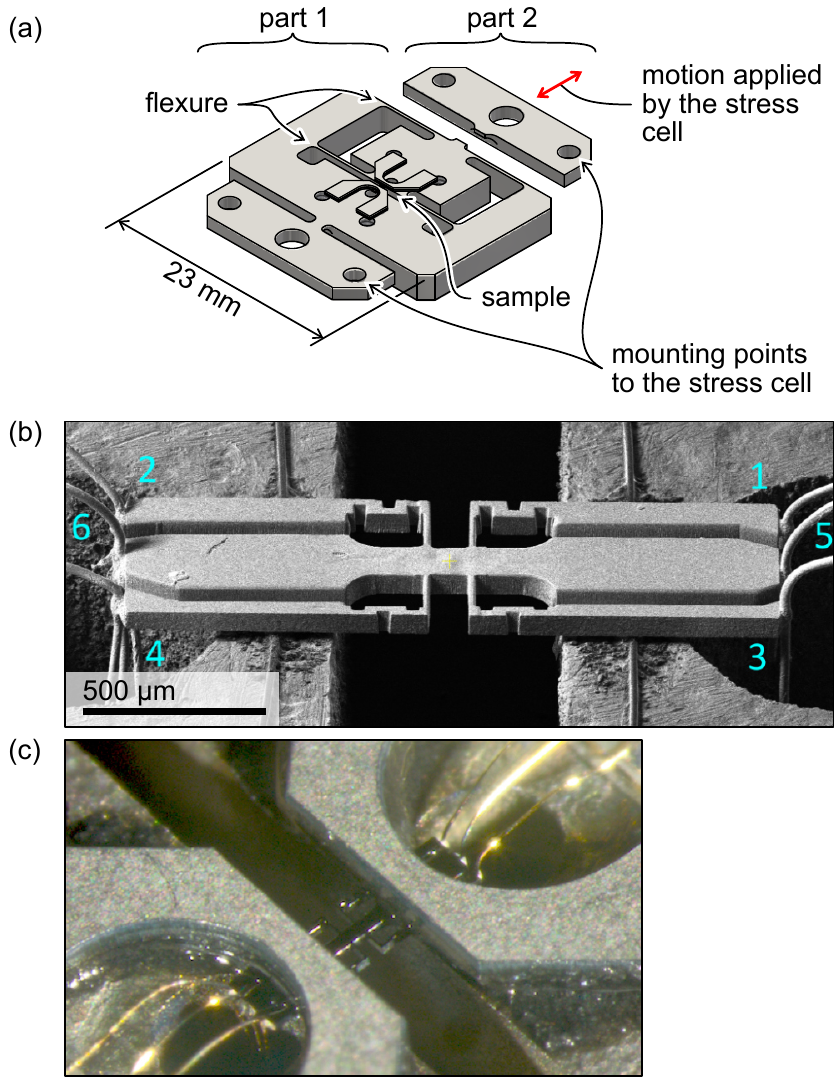}
\caption{(a) Illustration of the sample carrier used here. This carrier gets mounted to a uniaxial stress cell similar to that described in Ref.~\cite{Barber19_RSI}. (b) Scanning
electron micrograph of sample 2 before installation of the upper plates. (c) Photograph of sample 2 after installation of the upper plates.}
\end{figure}

\textit{The longitudinal resistivity at $\sigma = \sigma_\text{L}$.} In the main text, we used a relationship $[\rho_{xx}(B=0)/\Delta \rho_{xx}]^{0.5} = \alpha + \beta T^2$ to
subtract the magnetoresistance and estimate the $\rho_{xx}(T)$ that would be obtained at $B=0$ without superconductivity. We show in Fig.~S2 that the form $[\rho_{xx}(B=0)/\Delta
\rho_{xx}]^{0.5} = \alpha + \beta T^{1.5}$ can also be fit to the data at $\sigma = \sigma_\text{L}$. Potentially, it is a better model because non-umklapp scattering is expected
to be proportional to $T^{1.5}$ at $\sigma = \sigma_\text{L}$. Results when this model is used to subtract the magnetoresistivity are also shown in Fig.~S2. The conclusion is
the same as with the $T^2$ magnetoresistance model shown in the main text: $\rho_{xx} = \rho_0 + AT^{3/2}$ is a poor fit to the data, while $\rho_0 + A T^2 \log (T_0 / T)$ and
$\rho_0 + A T^{5/3}$ work equally well.
\\

\begin{figure}[ptb]
\includegraphics[width=85mm]{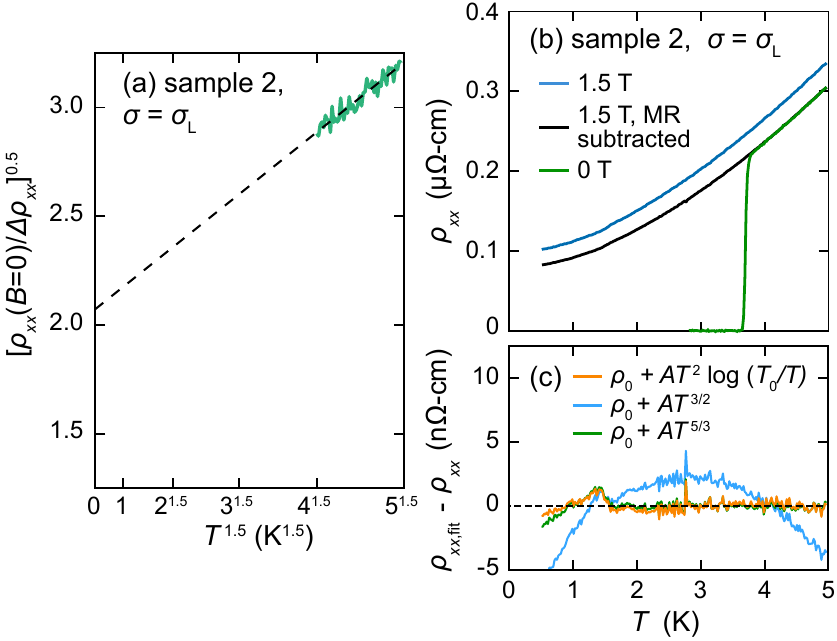}
\caption{An alternative magnetoresistance model. (a) $(\rho_{xx}(B=0) / \Delta \rho_{xx})^{0.5}$, where $\Delta \rho_{xx}$ is the change in $\rho_{xx}$ between 0 and 1.5~T, against
$T^{3/2}$ for sample 2 at $\sigma = \sigma_\text{L}$. (b) $\rho_{xx}(T)$ of sample 2 at $\sigma = \sigma_\text{L}$.  The black line is an average of the data at $+1.5$ and $-1.5$~T
with the magnetoresistance subtracted as described in the supplemental text. (c) Difference between three fitting models and the data with the magnetoresistance subtracted.}
\end{figure}

\textit{$V_\text{H}(B)$ for sample~1.} In Fig.~S3 we show the Hall voltage $V_\text{H}(B)$ for sample~1, at various temperatures. Due to its higher defect scattering, the low-field
range where $V_\text{H} \propto B$ extends slightly further than for sample~2.
\\

\begin{figure}[ptb]
\includegraphics[width=85mm]{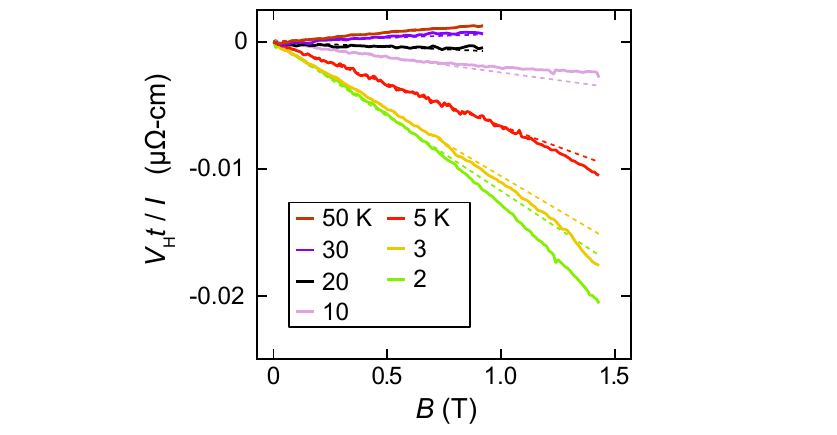}
\caption{Hall voltage $V_\text{H}$ versus field $B$ for sample~1 at various temperatures.}
\end{figure}

\textit{Complete data set at $B = \pm 0.35$~T.} In Fig.~4 in the main text, we excluded from panels (a) and (c), which show data at $B = \pm 0.35$~T, data that we estimated to be
affected by the superconductivity of \SRO{}. In Fig.~S4(a--b), we show the complete data set. 

In Fig.~S4(c) we also show an example of unsymmetrized data at 0.35~T: the voltage between contacts 1 and 3 divided by the applied current. We noted in the main text that the
stress-induced magnetic order has a strong effect on $R_\text{H}$ at 1.5~T, but no discernible effect on $R_\text{H}$ at 0.35~T. Here, it can be seen that there are anomalies in
the unsymmetrized data at 0.35~T. In Fig.~4(c) in the main text, there are also no clear anomalies in $\rho_{xx}(\sigma)$ when the magnetic order onsets, so the appearance of
anomalies in the unsymmetrized data suggests that the magnetic order affects $\rho_{yy}$ and/or $\rho_{zz}$ more strongly than $R_\text{H}$. $\rho_{yy}$ and $\rho_{zz}$ are
expected to enter the unsymmetrized Hall data at some level due to misalignment of the sample axes with respect to the crystal axes; without particular effort, misalignment of
$\sim 1^\circ$ is typical. The main point of showing this data is to make clear that the magnetic order is present at 0.35~T, as at 1.5~T.
\\

\begin{figure}[ptb]
\includegraphics[width=85mm]{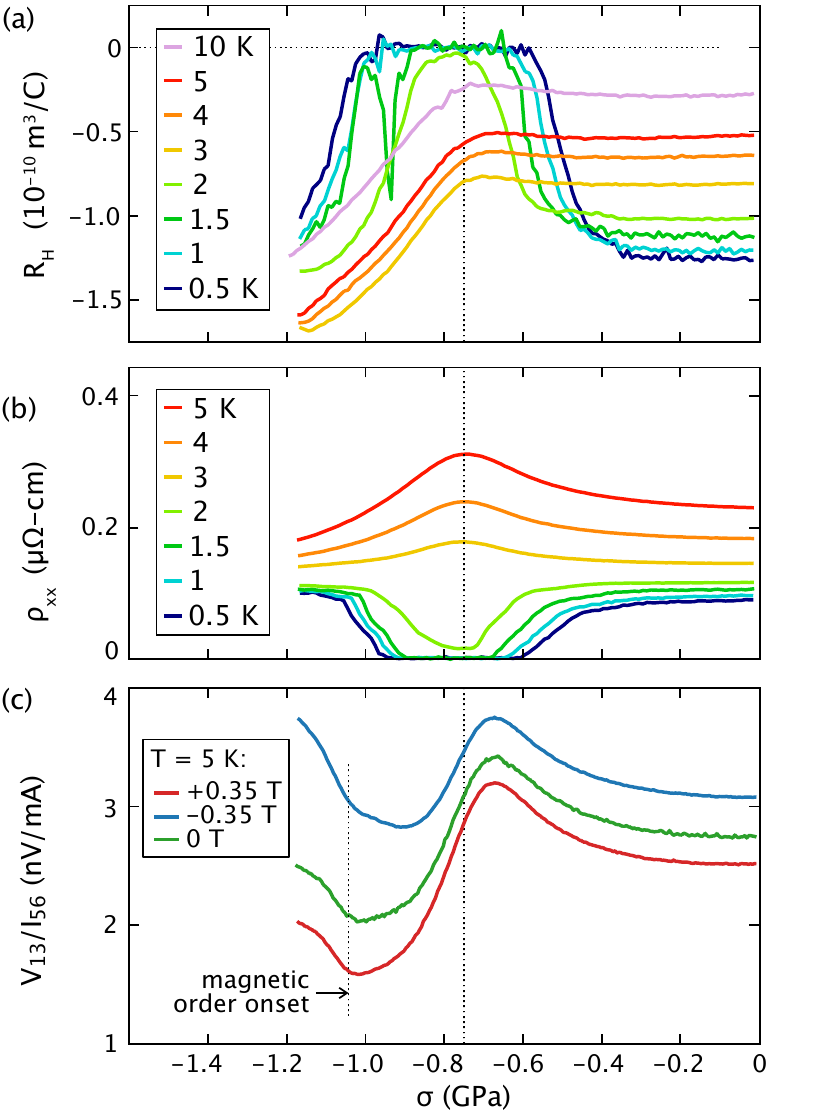}
\caption{(a) $R_\text{H}(\sigma)$ of sample 2, measured at $B = \pm 0.35$~T. This is a copy of Fig.~4(a) from the main text, but including the data that are affected by the
superconductivity. (b) $\rho_{xx}(\sigma)$ of sample~2 at $B = +0.35$~T--- Fig.~4(b) from the main text, with the data affected by the superconductivity included. (c) Unsymmetrized
data at $T = 5$~K and $\pm 0.35$~T: $V_{13}$, the voltage difference between contacts $1$ and $3$, divided by the current applied between contacts $5$ and $6$. The onset of the
magnetic order is taken as the stress where $d^2 R_\text{H}/d \sigma^2$ at 1.5~T is maximum; see Fig.~4(d).}
\end{figure}

\textit{Tight-binding model.} Our tight-binding model is:

\begin{eqnarray*}
H & = & \psi^\dagger_s(\mathbf{k})\left(\begin{matrix}E_{xz}(\mathbf{k}) & -isg & ig \\ isg & E_{yz}(\mathbf{k}) & -sg \\ -ig & -sg &
E_{xy}(\mathbf{k})\end{matrix}\right)\psi_s(\mathbf{k}),\\
E_{xz}(\mathbf{k}) & = & -\mu - 2t_1\cos k_x - 2t_2\cos k_y, \\
E_{xy}(\mathbf{k}) & = & -\mu - 2t_3(\cos k_x + \cos k_y) - 4t_4\cos k_x\cos k_y - \\
& & 2t_5(\cos 2k_x + \cos 2k_y).
\end{eqnarray*}
$\psi_s(\mathbf{k}) = [c_{\mathbf{k}, {xz}, s}, c_{\mathbf{k}, {yz}, s}, c_{\mathbf{k}, {xy}, -s}]^T$, $s = \pm 1$ for spin, and  $\{t_1, t_2, t_3, t_4, t_5, \mu, g\} = \{88, 9,
80, 40, 5, 109, 35\}$~meV. This model was derived as a fit to ARPES data in Ref.~\cite{Zabolotnyy2013_JoESaRP}, and was also applied in in Ref.~\cite{Romer2019_PRL, Cobo2016_PRB}.
These parameters yield cyclotron effective masses for the $\alpha$, $\beta$, and $\gamma$ sheets of 5.4, 4.8, and $16.7m_e$, respectively.

To apply uniaxial stress, $x$-oriented hopping integrals are scaled as $t \rightarrow t \times (1 - \alpha \varepsilon_{xx})$, and $y$-oriented hopping integrals as $t \rightarrow
t \times (1 + \alpha \nu_{xy}\varepsilon_{xx})$. $\varepsilon_{xx}$ is the longitudinal strain, $\nu_{xy} = 0.508$ is the Poisson's ratio~\cite{Ghosh21_NatPhys}, and $\alpha$ is a
constant that places the Lifshitz transition at $\sigma_\text{L} = -0.75$~GPa. The diagonal hopping integral, $t_4$, is scaled as $t_4 \rightarrow t_4 \times [1 -
\frac{\alpha}{2}(1-\nu_{xy})\varepsilon_{xx}]$. The Fermi surfaces resulting from this tight-binding model are shown in Fig.~1 in the main text. In Figs.~S5(a--c) we show these
Fermi surfaces again, but populated with arrows whose lengths are proportional to the mean free path. For this illustration, we take $\eta_{xy} = \eta_{xz, yz}$, so $l_\text{mfp}$
is proportional to the Fermi velocity. The Ong construction is illustrated in Figs.~S5(d--f). The curves shown in these panels are the paths traced when each mean free path vector
$\mathbf{l_\text{mfp}}$ is shifted to the origin. Ong showed that the Hall conductivity $\sigma_{xy}$ is proportional to the sum of the areas enclosed.

\begin{figure}[ptb]
\includegraphics[width=85mm]{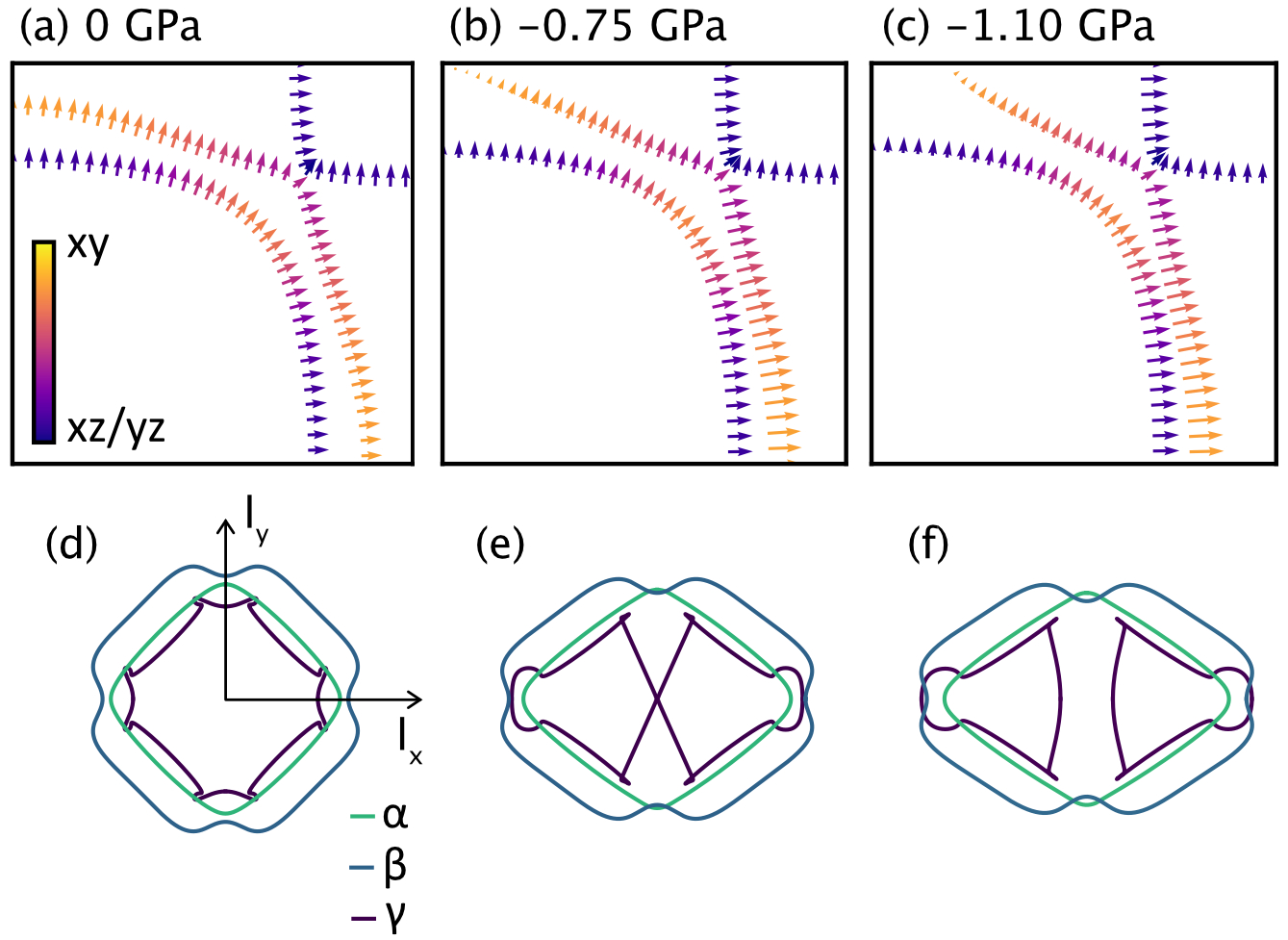}
\caption{(a--c) Fermi surfaces calculated using the tight-binding model described in the text. The arrow length is proportional to the Fermi velocity. (d--f) $\textbf{l}$ curves
under an assumption of isotropic scattering time, that is, $\eta_{xy} = \eta_{xz, yz}$. In this case, $\mathbf{l}_\text{mfp} \propto \mathbf{v}_\text{F}$.}
\end{figure}

\begin{acknowledgments}
\textit{Acknowledgements.} P.Y. thanks K. Shirer and M. K\"{o}nig for assistance with the plasma focused ion beam. We thank A. Georges, M. Zingl, and J. Mravlje for a critical read of the manuscript. We
additionally thank E. Berg, N. Hussey, and J. Schmalian for useful comments. We acknowledge the financial support of the Max Planck Society.  A.P.M. and C.W.H. acknowledge the
financial support of the Deutsche Forschungsgemeinschaft (DFG, German Research Foundation) - TRR 288 - 422213477 (project A10). N.K. is supported by a KAKENHI Grants-in-Aids for
Scientific Research (Grant Nos. 17H06136, 18K04715, and 21H01033), and Core-to-Core Program (No.  JPJSCCA20170002) from the Japan Society for the Promotion of Science (JSPS) and by
a JST-Mirai Program (Grant No. JPMJMI18A3). H. M. L. N. acknowledges support from the Alexander von Humboldt Foundation through a Research Fellowship for Postdoctoral Researchers.
Research in Dresden benefits from the environment provided by the DFG Cluster of Excellence ct.qmat (EXC 2147, project ID 390858940).
Raw data for this publication are available at \textit{url to be determined.}
\end{acknowledgments}

\bibliography{bibliography_SRO.bib}

\end{document}